\documentclass[11pt,a4paper]{article}
\pdfoutput=1

\usepackage{jheppub}
\usepackage[english]{babel}
\usepackage[utf8]{inputenc}
\usepackage{epsf}
\usepackage{graphicx}
\usepackage{subcaption} % allows subfigures

\usepackage{verbatim} % allows block comments

\usepackage{amsmath}
\usepackage{amssymb}
\usepackage{mathrsfs}

\usepackage{mathbbol}

\usepackage{amsfonts}
\usepackage{bbding}
\usepackage{array}
\usepackage{lineno}

\usepackage{color} % colored text
\usepackage{tikz}
\usetikzlibrary{shapes,arrows,positioning,automata,backgrounds,calc,er,patterns}
\usepackage[compat=1.1.0]{tikz-feynman}
\vspace*{1cm}
\allowdisplaybreaks
\title{Constraints on asymmetric production of long-lived scalars at the Large Hadron Collider}

\author[1]{Thomas Chehab}
\author[2]{\!\!, Louie Dartmoor Corpe}
\author[2]{\!\!, Andreas Goudelis}
\author[2]{\!\!, Abdelhamid Haddad}
\author[3]{\!\!, Louise Millot}

\affiliation[1]{Artemis, Université Côte d’Azur, CNRS, Observatoire Côte d’Azur, BP4229, 06304, Nice Cedex 4, France}
\affiliation[2]{Laboratoire de Physique de Clermont Auvergne (UMR 6533), CNRS/IN2P3, Univ.\ Clermont Auvergne, 4~Av.\ Blaise Pascal, F-63178 Aubi\`ere Cedex, France}
\affiliation[3]{Laboratoire de Physique Subatomique et de Cosmologie (LPSC), Universit\'e Grenoble-Alpes, CNRS/ IN2P3, 53 Avenue des Martyrs, F-38026 Grenoble, France}

\abstract{
Searches for pair-produced long-lived particles (LLPs) at the LHC commonly operate under the assumption that the two LLPs are identical. In this paper we entertain the possibility that the targeted final states are, instead, induced by a LLP pair with different masses and/or lifetimes. We propose a simple and intuitively-parametrised toy model in order to study such asymmetric production of LLPs. Using the recasting material of a recent search for displaced jets by the ATLAS collaboration, we demonstrate that we can set constraints on the production cross-section times branching fraction into jets for a variety of asymmetric LLP and mediator mass combinations.
}
\begin{document}
\maketitle

%%%%%%%%%%%%%%%%%%%%%%%%%%%%%%%%%%%%%%%%%%%%%%%%%%%%%%%%%%%%%%%%%%%%%%%%%%%%%%%%%%%
%%%%%%%%%%%%%%%%%%%%%%%%%%%%%%%%%%%%%%%%%%%%%%%%%%%%%%%%%%%%%%%%%%%%%%%%%%%%%%%%%%%
%%%%%%%%%%%%%%%%%%%%%%%%%%%%%%%%%%%%%%%%%%%%%%%%%%%%%%%%%%%%%%%%%%%%%%%%%%%%%%%%%%%
\section{Introduction}\label{sec:intro}

During the past decade, searches for new long-lived particles (LLPs), \textit{i.e.} particles with a macroscopic lifetime, have evolved from a relatively ``exotic'' topic into an important pillar of the LHC New Physics search programme -- and for good reasons.
Indeed, LLPs not only represent an exciting experimental challenge, among other reasons because they touch upon the very definition of what constitutes a ``particle'' from an experimental standpoint\footnote{As an example, and grossly speaking, experimentally a promptly produced electron is a track accompanied by energy deposits in the electromagnetic calorimeter. An electron stemming from a displaced vertex, on the other hand, may be a trackless object. From a theoretical perspective, of course, both objects are identical.}, but are also theoretically well-motivated. Examples of classes of new physics scenarios potentially predicting the existence of exotic LLPs include Hidden Valley models~\cite{Strassler:2006im,Strassler:2006ri,Chan:2011aa}, models of feebly interacting dark matter~\cite{Calibbi:2018fqf,Belanger:2018sti}, neutrino mass models~\cite{Minkowski:1977sc,Glashow:1979nm} or supersymmetric~\cite{Arkani-Hamed:2012fhg,Giudice:1998bp,Barbier:2004ez,Csaki:2015fea,Fan:2011yu,Fan:2012jf} and/or ``neutral naturalness'' scenarios~\cite{Chacko:2015fbc,Burdman:2006tz,Cai:2008au,Chacko:2005pe}.

Given that so far no significant excesses have been observed, and although variations depending on the specific final state being considered may exist, typical LLP searches present their results in the form of constraints on the LLP production cross-section times branching fraction ($\sigma \times {\rm BF}$) as a function of the LLP lifetime and mass. To do so, analyses often use simple benchmark models which, essentially, are used as kinematic templates to guide the optimisation of the analyses -- \textit{i.e.} to define a set of kinematic requirements for signal selection and background rejection. It follows that the limits (published in terms of physical quantities such as cross-sections, masses, and lifetimes) only hold within the framework of the underlying template model, or within very contrived embeddings in larger frameworks that contain the template model in essentially intact/factorisable form. 

In this paper we focus on the case of ATLAS and CMS searches for neutral, hadronically-decaying scalar LLPs (see Refs.~\cite{CMS-EXO-19-011, HDMI, ATLAS-EXOT-2019-23,EXOT-2017-05}
for recent examples). These searches are designed using the Hidden Abelian Higgs Model (HAHM) \cite{Wells:2008xg} as a template, assuming that a pair of identical scalar LLPs are produced through the decay of a scalar mediator and subsequently decay, through Yukawa-like interactions, predominantly into heavy quarks (\textit{i.e.} jets) and/or $\tau^+ \tau^-$ pairs. The results are, then, presented in terms of constraints on the mediator production cross-section times branching ratio into LLP pairs decaying into jets.

There are at least three instances, or combinations thereof, in which it would be natural to expect a substantial modification of the published constraints:
\begin{itemize}
    \item Multiple states (\textit{e.g.} Higgs bosons) could mediate the production of LLPs, which, if their contributions are comparable, could alter the kinematic distributions of the decay products;
    \item Gluon fusion might not be the predominant mediator production mode, and jets might not be the LLP decay mode driving experimental sensitivity (\textit{e.g.} in models involving vector LLPs, or, in the case of scalars, Yukawa coupling structures substantially departing from a Minimal Flavour Violation-like hypothesis);
    \item The produced LLPs could, simply, \textit{not} be identical. From a phenomenological standpoint, this would mean that their mass and/or mean proper lifetime could be different which would then lead to modifications in the decay products' kinematic and/or spatial distributions. Such a situation could arise, for example, in models involving extended scalar sectors with multiple singlets coupled to the Standard Model (SM) Higgs boson bilinear $H^\dagger H$.
\end{itemize}
In this paper we focus on the latter possibility, which we will henceforth refer to as ``asymmetric'' LLP production. Our analysis will be mainly driven by phenomenological considerations: we aim neither at considering the ``best-motivated'' SM extension nor at studying a model which can encapsulate all New Physics scenarios predicting the existence of multiple LLPs. Instead, we will study how the constraints presented in Ref.~\cite{ATLAS-EXOT-2022-04} translate in terms of a simple and, hopefully, intuitively parametrised toy model of ``asymmetrically produced'' LLPs, depending on their masses and lifetimes. 

The paper is structured as follows. In Sec.~\ref{sec:habits} we describe the ATLAS search for LLPs which we will use as a reference throughout the paper, as well as the HAHM which it uses as a benchmark model. In Sec.~\ref{sec:model} we propose an alternative lifetime-parametrised model, which also has the flexibility to allow asymmetric LLP production. This is followed by Sec.~\ref{sec:results} where we use the recasting material for the reference search to derive, for the first time, limits on asymmetric LLP production. Finally, we draw our conclusions in Sec.~\ref{sec:conclusions}.

%%%%%%%%%%%%%%%%%%%%%%%%%%%%%%%%%%%%%%%%%%%%%%%%%%%%%%%%%%%%%%%%%%%%%%%%%%%%%%%%%%%
%%%%%%%%%%%%%%%%%%%%%%%%%%%%%%%%%%%%%%%%%%%%%%%%%%%%%%%%%%%%%%%%%%%%%%%%%%%%%%%%%%%
%%%%%%%%%%%%%%%%%%%%%%%%%%%%%%%%%%%%%%%%%%%%%%%%%%%%%%%%%%%%%%%%%%%%%%%%%%%%%%%%%%%
\section{Reference search for displaced jets with the ATLAS detector}\label{sec:habits}

The analysis presented in Ref.~\cite{ATLAS-EXOT-2022-04} targets pairs of identical neutral long-lived particles decaying hadronically in the ATLAS hadronic calorimeter system. It is part of a family of searches which look for the same signature with decays occurring in various parts of the ATLAS detector~\cite{HDMI,ATLAS-EXOT-2019-23,EXOT-2017-05}. Each of those searches, and their various iterations, have used the HAHM~\cite{Wells:2008xg} (which will be briefly discussed in Sec~\ref{sec:hahm}) as a kinematic template since the start of LHC Run 2.  In this paper, we use the most recent calorimeter-based ATLAS search as a reference because it has the advantage of providing some of the most-cutting edge re-interpretation material available~\cite{Zenodo}: a machine-learning based ``surrogate model'' to obtain the probability that an event would be selected based only on truth-level inputs, \textit{i.e.} circumventing the need for detector simulation. An independent validation of the surrogate models for that analysis is available in Ref.~\cite{Corpe:2025sbw}. The availability of this recasting material is particularly relevant because many efforts to recast LLP analyses cannot reliably make use of tools like \textsc{Delphes}~\cite{deFavereau:2013fsa} to simulate the effects of the detector since most LLP searches rely heavily on the detailed response of each layer. For this reason, it is often difficult to estimate the sensitivity of such a search to a new model, even when it is relatively close to the benchmark models employed in the original analysis. The presence of the surrogate model in the re-interpretation material is a boon for this type of exercise.

%%%%%%%%%%%%%%%%%%%%%%%%%%%%%%%%%%%%%%%%%%%%%%%%%%%%%%%%%%%%%%%%%%%%%%%%%%%%%%%%%%%
\subsection{Description of reference analysis}

We will now briefly describe the reference analysis~\cite{ATLAS-EXOT-2022-04}, which is an extension of a similar analysis presented in Ref.~\cite{ATLAS-EXOT-2019-23}. In both cases, the full LHC Run 2 dataset is used to search for pairs of displaced jets originating from hadronic decays of neutral LLPs in the hadronic part of the calorimeter. 
The displaced jets are expected to be largely trackless (since the LLP is neutral) and with a high proportion of energy in the hadronic part of the calorimeter with respect to the upstream electromagnetic calorimeter (since the search focuses on decays in the hadronic part). These selections are helpful to reduce  QCD multijets, which only rarely produce such signatures, but which still make up the dominant background on account of their overwhelming production cross-section in proton collisions. The data were collected with dedicated triggers which also exploit this calorimeter energy ratio topology. Both analyses makes use of a neural network (NN), whose inputs include deposits in each layer of the calorimeter as well as tracker and muon system information. It classifies jets as likely to have originated from QCD processes, signal, or beam-induced background (such as beam halo muons which travel parallel to the LHC beam and strike the detector without passing through the interaction point). 

The search presented in Ref. \cite{ATLAS-EXOT-2019-23} was optimized for cases where the decay products of each LLP (for instance two $b$-quarks) are very collimated, leading to a single ``merged'' jet in each case. This is optimal for cases where the mediator is heavy, but breaks down when the mediator is relatively light, because a large fraction of LLP decays can lead to two resolvable jets. Hence, the analysis was re-optimised in Ref. \cite{ATLAS-EXOT-2022-04}, to allow one of the jets to be ``resolved''.   The analysis also contained channels where the mediator was produced in association with a $W$ or $Z$ boson, but we do not consider those channels further in this work: we focus, instead, exclusively on the ``CR+2J'' channel (defined in~\cite{ATLAS-EXOT-2022-04}), optimised for cases where the mediator is produced via gluon fusion.  This channel  makes use of event-level machine-learning to exploit the NN scores for the jets most likely to originate from a signal process, along with additional event-level information, to classify events as signal-like or not. Some additional cleaning requirements (on signal-like jet timing, momentum and position, etc...)  are used to remove any remaining events likely to have originated from cosmic ray muons or beam-induced background. 
The resulting event-level score is then used with an orthogonal variable measuring the proximity of tracks to the jets in the events, to form an ABCD plane with which a data-driven background estimate can be made.  The signal is mostly concentrated in region A of the plane, while the background (which at this stage is composed purely of QCD multijets) is distributed in the other regions in an uncorrelated way. 
A simultaneous fit of the signal and background yields (which should obey the $A=B\times C/D$ relation) is performed for the final statistical analysis. 
The signal efficiencies in region A are extrapolated to arbitrary lifetimes using an exponential reweighting method. We note that this method, adopted for reasons of computational efficiency (\textit{i.e.} to minimize Monte Carlo event generation and detector simulation) clearly assumes that the mean proper lifetime is uncorrelated with the rest of the model parameters. This, of course, is not true and will be further commented upon in what follows.
In the absence of an excess, the CL$_s$~\cite{Read:2002hq} method was used to set constraints on $\sigma \times \textrm{BF}$ as a function of the LLP mass, for a set of benchmark mediator masses. As mentioned above, the surrogate models provided by the analysis and validated in~\cite{Corpe:2025sbw}, provide an easy-to-use proxy to the analysis selection efficiency for scenarios with different kinematics than the original template model, such as the one we study in this paper. 
%%%%%%%%%%%%%%%%%%%%%%%%%%%%%%%%%%%%%%%%%%%%%%%%%%%%%%%%%%%%%%%%%%%%%%%%%%%%%%%%%%%
\subsection{Underlying assumptions: the HAHM  in theory and in practice}\label{sec:hahm}

As we already mentioned, the microscopic model that was chosen as a kinematic template in Refs.~\cite{ATLAS-EXOT-2019-23, ATLAS-EXOT-2022-04} was the HAHM~\cite{Wells:2008xg}. The HAHM extends the SM with an additional $U(1)_X$ gauge group under which all SM particles are neutral and which is spontaneously broken by the vacuum expectation value of a $SU(3)_c \times SU(2)_L \times U(1)_Y$-singlet scalar field. The model predicts the existence of an additional gauge boson ($Z'$) along with a new scalar particle ($S$) which mixes with the SM Higgs boson. This mixing, in turn, induces couplings of the new state to the SM fermions. 

Focusing on the scalar sector, the concrete process through which displaced jets can in principle arise in the HAHM (albeit with an extremely small cross-section) is gluon fusion-induced (potentially virtual) SM Higgs boson ($h$) production, followed by the production of two exotic scalar states $S$, which in this case plays the role of the LLP. These  can, in turn, decay into SM particles with a macroscopic lifetime, provided the mixing with the SM Higgs boson is small enough. It should be noted that, once all of the HAHM free parameters are fixed, the LLP mean proper lifetime (along with all the cross-sections, partial widths and branching ratios, for that matter) is a \textit{predicted} quantity.

\begin{figure}[h!]
    \centering
    \includegraphics[width=0.75\linewidth]{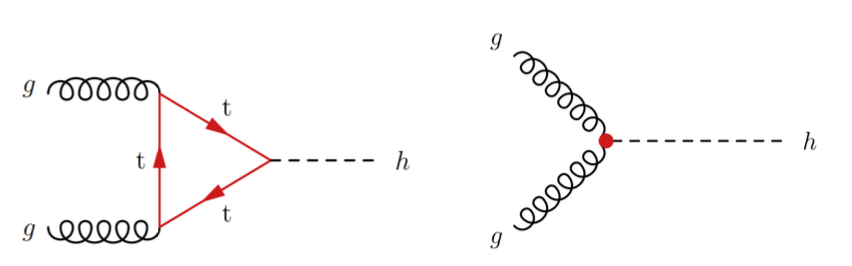}
    \caption{Gluon fusion-induced mediator production in the HAHM in the dynamical (left) and infinite mass (right) top quark limit.}
    \label{fig:hahm-ggH-effective-vertex}
\end{figure}

In the analyses presented in Refs.~\cite{ATLAS-EXOT-2019-23, ATLAS-EXOT-2022-04}, the implementation of the HAHM in MadGraph~\cite{Alwall:2014hca} presented in Ref.~\cite{Curtin:2014cca} was employed, where gluon fusion was approximated by an effective interaction of the SM Higgs boson with a pair of gluons, based on the prediction obtained in the heavy top quark loop limit, as illustrated in Figure~\ref{fig:hahm-ggH-effective-vertex}.
\\
\\
Using this HAHM implementation, the event generation in Refs.~\cite{ATLAS-EXOT-2019-23, ATLAS-EXOT-2022-04} proceeds as follows: 
\begin{itemize}
    \item Events are generated for chosen combinations of $m_{h,S}$, the masses of the two scalars, which are treated as free parameters. In other words, the SM Higgs Boson is co-opted into the role of a generic, gluon fusion-induced scalar mediator. In such cases, we denote the particle instead by $\Phi$\footnote{It follows that, when events are generated with a mediator mass assumption other than 125 GeV, there is actually no SM-like Higgs boson in the model; in this instance, the HAHM is just meant to be used as a template for kinematics.}, whereas the role of LLPs is undertaken by the exotic scalars;
    \item For each chosen set of $m_{h,S}$ values, a single event sample is actually generated, with a mean proper lifetime \textit{chosen} to maximise the number of observable decays in the calorimeters, regardless of the values of the underlying parameters. To be even more explicit: with the typical mass and coupling values actually chosen for the initial event generation in the ATLAS searches, it turns out that the HAHM actually predicts \emph{promptly} decaying exotic neutral scalars. The macroscopic lifetime is imposed \textit{post-hoc}, by replacing the decay position in the Les Houches Events file by hand.
\end{itemize}
These comments motivate the following considerations. The HAHM is a well-established, fully consistent (\textit{e.g.} gauge-invariant and renormalizable) model of New Physics. However \textit{i)} its usage in practice introduces a set of \textit{phenomenological} inconsistencies described in the bullets above and \textit{ii)} these inconsistencies are opaque both to the analysis team and to the community at large. Could an alternative simplified model be considered which, in a sense, actually ``accepts its nature'' as serving as a mere kinematic template and which manages to address at least some of these concerns?  This would perhaps come at the expense of theoretical consistency, but at least would provide some more insight into the event generation procedure, in the sense of being parametrised in terms of more physically intuitive quantities such as the LLP lifetime itself.

Clearly, the answer to this question is neither easy, nor universal. In what follows we will propose such a framework for our case-study analysis, being nonetheless fully conscious of the fact that there might be cases in which it may \textit{not} be possible to employ ``toy models'' which are as simple\footnote{One such example are, \textit{e.g.}, searches which are performed having in mind processes involving multiple gauge bosons, in which gauge invariance can become important for Monte Carlo event generation.}.

%%%%%%%%%%%%%%%%%%%%%%%%%%%%%%%%%%%%%%%%%%%%%%%%%%%%%%%%%%%%%%%%%%%%%%%%%%%%%%%%%%%
%%%%%%%%%%%%%%%%%%%%%%%%%%%%%%%%%%%%%%%%%%%%%%%%%%%%%%%%%%%%%%%%%%%%%%%%%%%%%%%%%%%
%%%%%%%%%%%%%%%%%%%%%%%%%%%%%%%%%%%%%%%%%%%%%%%%%%%%%%%%%%%%%%%%%%%%%%%%%%%%%%%%%%%
\section{A simple approach for asymmetric production of LLPs}\label{sec:model}

Let us now turn to our main question, \textit{i.e.} how the constraints obtained in Refs.~\cite{ATLAS-EXOT-2019-23, ATLAS-EXOT-2022-04} translate in the case in which a gluon-fusion-produced scalar mediator decays into a pair of different scalar LLPs. In order to describe the process, we consider the Lagrangian
\begin{align}\label{eq:TheLagrangian}
    {\cal{L}} & = {\cal{L}}_{f} + {\cal{L}}_{\rm QCD} -\frac{m_\Phi^2}{2} \Phi^2 - \frac{m_{S_1}^2}{2} S_1^2 - \frac{m_{S_2}^2}{2} S_2^2  \\ \nonumber
& + \frac{1}{\Lambda} \Phi G_{\mu\nu} G^{\mu\nu} + \kappa v \Phi S_1 S_2 + \sum_{f} \left( y_1 \frac{\sqrt{2} m_f}{v} S_1 f\bar{f} + y_2 \frac{\sqrt{2} m_f}{v} S_2 f\bar{f} \right)
\end{align}
where ${\cal{L}}_{f}$ and ${\cal{L}}_{\rm QCD}$ denote the SM fermion gauge interaction and QCD Lagrangians, respectively, $\Phi$ is the mediator, $G$ is the gluon field strength tensor, $\Lambda$ is a scale governing gluon fusion-induced mediator production, $S_{1,2}$ are the LLPs, $\kappa$ is a dimensionless coefficient describing the interaction strength between the LLPs and the mediator, $y_{1,2}$ are the LLP Yukawa-like couplings to the SM fermions $f$ and $v$ is the SM Higgs boson vacuum expectation value. The normalisation factors have been chosen with a ``doublet-singlet'' model (\`a la HAHM) in mind, but the exact choice is irrelevant (\textit{i.e.} it can be compensated by a corresponding rescaling of the different coupling constants). The last two terms of Eq. \eqref{eq:TheLagrangian} do, however, imply that the LLPs couple to the SM fermions proportionally to their mass.
\\
\\
Now, in this framework the mediator and LLP partial decay widths acquire a simple form
\begin{align}
    \Gamma_\Phi^{jj} & = \frac{2 m_\Phi^3}{\pi \Lambda^2} \\
    \Gamma_\Phi^{\rm LLP} & = \frac{\kappa^2 v^2 \sqrt{\lambda(m_\Phi^2, m_{S_{1}}^2, m_{S_{2}}^2)}}{16\pi m_\Phi^3} 
\end{align}
and
\begin{equation}
    \Gamma_{S_i}^{ff} = \sum_{f} \left( \frac{C_f y_i^2 m_f^2 (m_{S_i}^2 - 4 m_f^2)^{3/2}}{\pi v^2 m_{S_i}^2} \right)\\
\end{equation}
where $C_f = 1(3)$ for leptons (quarks). The latter of these expressions (in which the sum runs over all SM fermions) can be easily inverted in order to obtain the overall couplings $y_i$ as a function of the LLP lifetime which, in turn, can be promoted to an input parameter of the model. The ``symmetric'' scenario can, then, be recovered simply by choosing the two masses and mean proper lifetimes to be the same.

All in all, the model described by the Lagrangian \eqref{eq:TheLagrangian} is governed by seven free parameters which we choose to be: the mediator and LLP masses, the two LLP lifetimes and the parameters $\Lambda$ and $\kappa$ which describe the mediator interactions with gluons (for $\Lambda$)  and $S_{1,2}$ (for $\kappa$), respectively\footnote{Note that in principle the latter two parameters can also be traded for the mediator partial widths into gluon and LLP pairs, respectively. Each choice offers different advantages. For example, promoting the partial widths to input parameters would make it easier to ensure that the mediator width remains smaller than its mass. Retaining the original Lagrangian parameters, on the other hand, allows for an easier assessment concerning the EFT scale $\Lambda$ value that is being probed.}.

It is clear that from a theoretical standpoint, the Lagrangian \eqref{eq:TheLagrangian} is not fully consistent. Concretely, it is not gauge-invariant. However, we believe that it nevertheless has the following practical advantages. 
\begin{itemize}
    \item It focuses on the particle content which is, in practice, relevant for the experimental search at hand. For example, the $Z'$ part of the HAHM has been discarded: searches for long-lived gauge bosons are, of course, extremely interesting but would most likely be optimally performed through final states involving leptons. Moreover, even fully hadronically decaying gauge bosons would typically be produced through $q\bar{q}$ annihilation which would, in turn, imply very different kinematics compared with gluon fusion-produced objects.
    \item It is intuitive in the sense that, at least as far as the LLP properties are concerned, it can be easily parametrised in terms of physically measurable quantities (masses, lifetimes). We consider this to be an important point, since it can allow experimental groups to better control the event generation process and, \textit{e.g.}, generate initial event samples which are as ``phenomenologically consistent'' as possible.
\end{itemize}
In order to better understand the second comment, let us give a concrete example: LLP decay widths are, necessarily, extremely narrow. The $S_1$ or $S_2$ width can, in principle, have an impact on the decay products' kinematic distributions and, hence, on the ensuing constraints. Even if, for reasons of computational efficiency, only a single event sample per $m_{\Phi,S_1, S_2}$ combination is generated, through the parametrisation that we propose it is completely straightforward to choose macroscopic $S_1$ or $S_2$ lifetimes and the associated decay widths will automatically and consistently be adjusted to be narrow. Let us also be clear on two points: first, as long as a single event sample is generated for each $m_{\Phi,S_1, S_2}$ combination, some amount of inconsistency is bound to persist; however, and at least, all constraints will be obtained under the assumption of a very narrow LLP width. Secondly, in practice we have found that, at least in this instance, the LLP width can vary over orders of magnitude without affecting the resulting constraints. In some sense this is expected, as long as $\Gamma_{S_i}^{ff}$ remains below the experimental energy resolution. However, the validity of this statement depends on one hand on the underlying model and, on the other hand, on the parameter values employed for the initial event generation. 

This, in our view, is one of the major advantages of the framework that we propose: the user can choose from the start the LLP lifetime as an input parameter and the LLP width is modified accordingly and consistently, at least for the original event generation sample.
%%%%%%%%%%%%%%%%%%%%%%%%%%%%%%%%%%%%%%%%%%%%%%%%%%%%%%%%%%%%%%%%%%%%%%%%%%%%%%%%%%%
%%%%%%%%%%%%%%%%%%%%%%%%%%%%%%%%%%%%%%%%%%%%%%%%%%%%%%%%%%%%%%%%%%%%%%%%%%%%%%%%%%%
%%%%%%%%%%%%%%%%%%%%%%%%%%%%%%%%%%%%%%%%%%%%%%%%%%%%%%%%%%%%%%%%%%%%%%%%%%%%%%%%%%%
\section{Numerical results}\label{sec:results}

Let us now turn to our numerical analysis. In order to generate events, we have implemented the model described by the Lagrangian \eqref{eq:TheLagrangian} in {\tt MadGraph5$\_$aMC@NLO} (MG) \cite{Alwall:2014hca} by means of the {\tt FeynRules} package~\cite{Alloul:2013bka}. Throughout the subsequent analysis we will be employing the {\tt NNPDF2.3Lo} set of Parton Distribution Functions whereas showering and hadronization are simulated with {\tt Pythia 8}~\cite{Bierlich:2022pfr} with all settings chosen at their default values.

%%%%%%%%%%%%%%%%%%%%%%%%%%%%%%%%%%%%%%%%%%%%%%%%%%%%%%%%%%%%%%%%%%%%%%%%%%%%%%%%%%%
\subsection{Validation: symmetric production of LLPs}\label{sec:validation}

The first step is to validate the theoretical and computational framework that we propose. Concretely, in this context the validation involves two distinct tasks:
\begin{itemize}
    \item First, we must ensure that our template model yields the same results as the HAHM in the ``symmetric'' limit, \textit{i.e.} when the two LLPs are assigned identical masses and lifetimes.
    \item Second, for reasons of computational efficiency in what follows, we will be adopting the approach followed by the ATLAS collaboration: generating one event sample per $(m_\Phi, m_{S_1}, m_{S_2})$ combination and sampling the LLP proper lifetime from an exponential distribution, regardless of the predicted LLP (and mediator) width. Thus, we need to establish that, at least within the scope of our analysis, both approaches yield quantitatively similar results.
\end{itemize}

\begin{figure}
    \centering
    \includegraphics[width=0.45\linewidth]{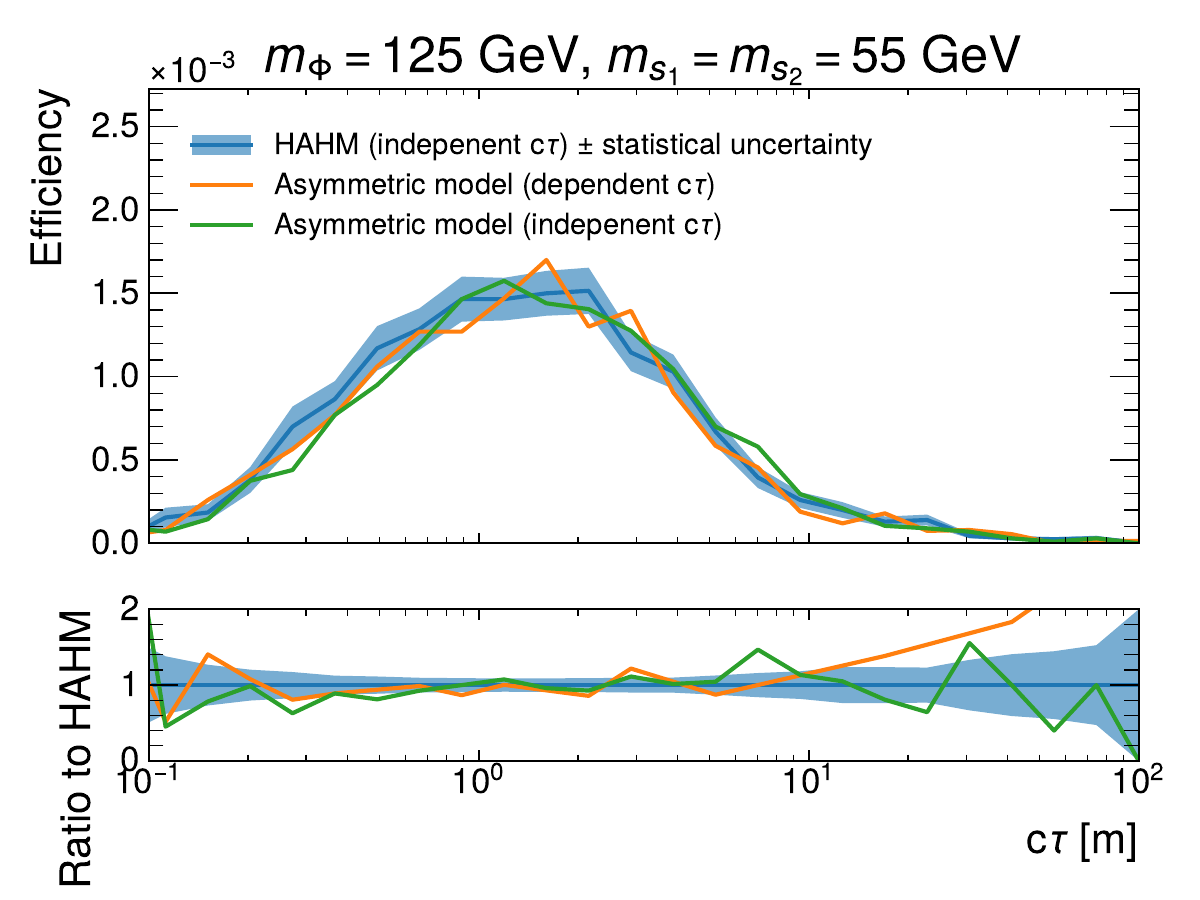}
   \includegraphics[width=0.45\linewidth]{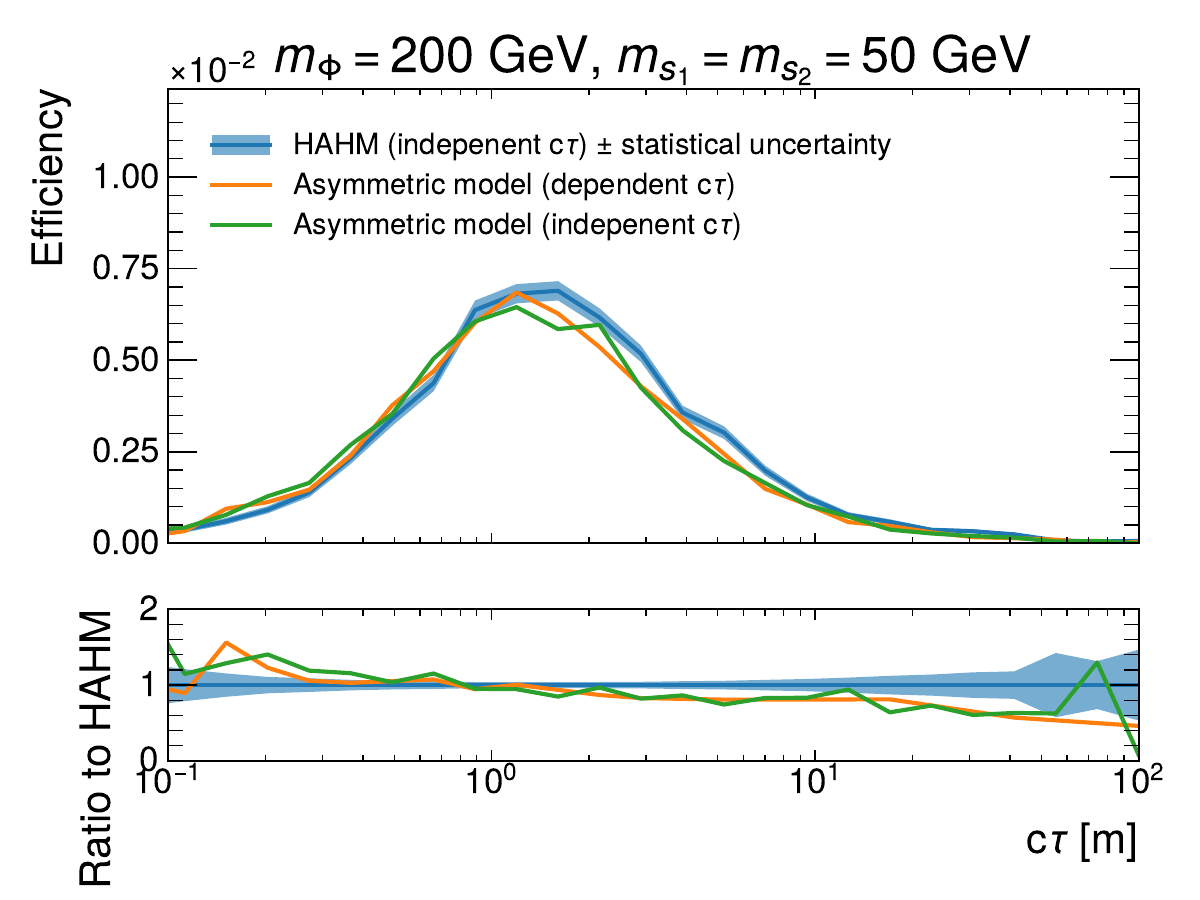}\\
      \includegraphics[width=0.45\linewidth]{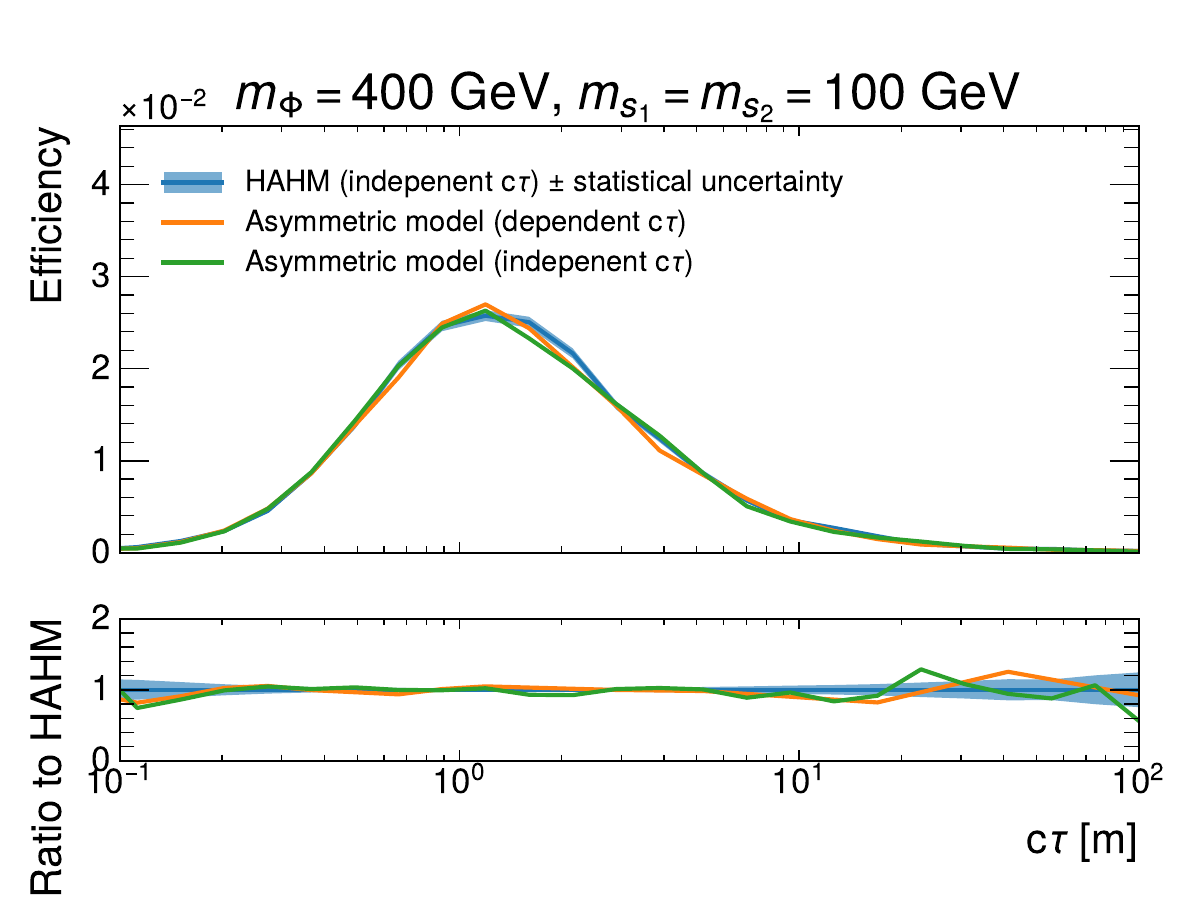}
   \includegraphics[width=0.45\linewidth]{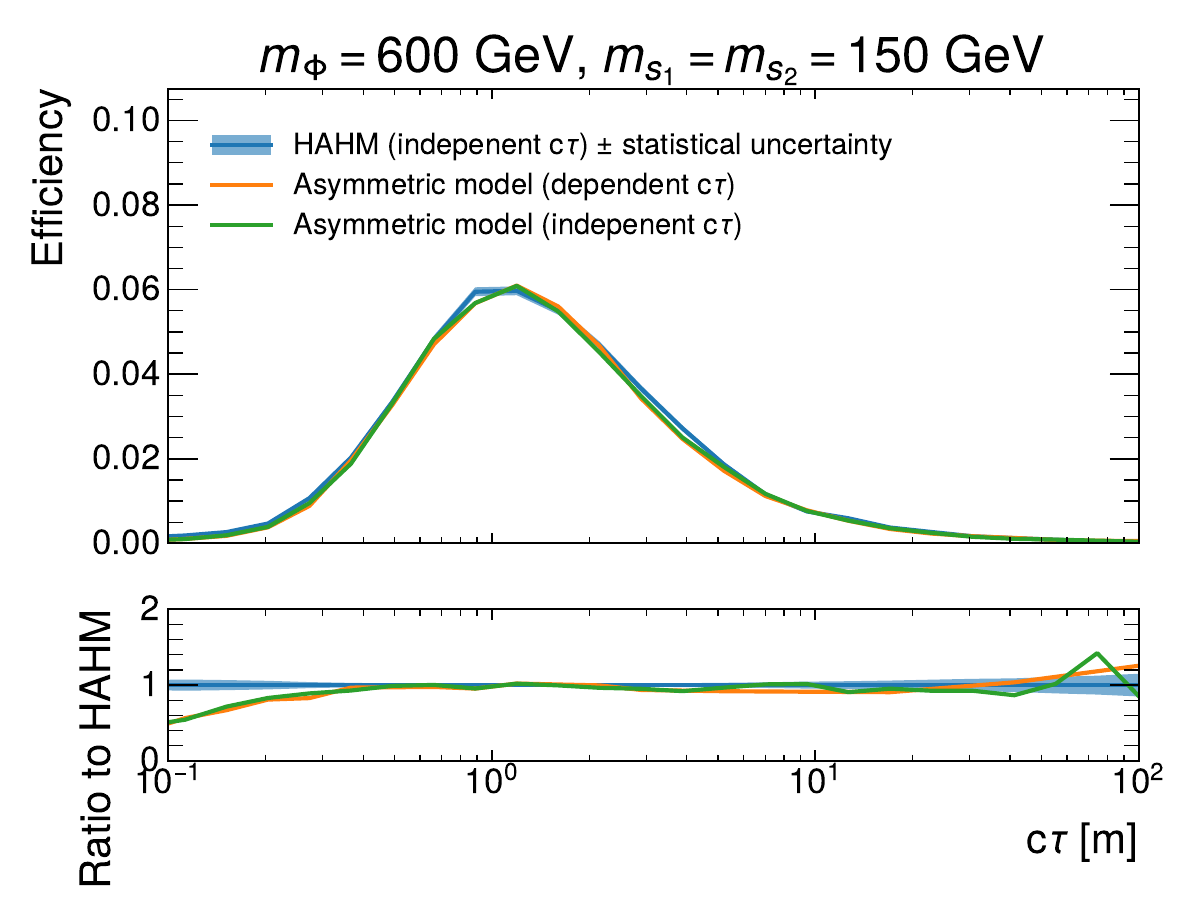}\\
  \caption{Validation plots comparing the selection efficiency obtained from the HAHM model kinematics, and the asymmetric model while treating the lifetime as a dependent or independent parameter.}
    \label{fig:lifetime-param-validation}
\end{figure}

In order to perform these validation exercises we first choose four benchmarks sets of masses (with $S_1$ and $S_2$ masses and lifetimes set to be equal), and we use the surrogate models described in Sec.~\ref{sec:habits} to obtain efficiencies for the events to be selected in Region A of the reference analysis. In Fig.~\ref{fig:lifetime-param-validation} we present the obtained efficiencies as a function of lifetime, focusing on the relative changes between \textit{(i)} the raw HAHM (labeled ``HAHM (independent $c\tau$)''), as used in the original ATLAS analysis, \textit{(ii)} the symmetric limit of the model presented in Sec.~\ref{sec:model} with a consistent treatment of the LLP lifetime, \textit{i.e.} maintaining all relations between $c\tau$ and the other parameters of the model (labeled ``asymmetric model (dependent $c\tau$)'') and \textit{(iii)} the same model, but instead treating the lifetime as an independent parameter by sampling from an exponential distribution  (labeled ``asymmetric model (independent $c\tau$)'').

Our results show that for the lifetime range in which the ATLAS searches are sensitive, the three approaches yield consistent results. These findings confirm two key points. First, the asymmetric model correctly reproduces the original HAHM when the two LLPs have identical masses and lifetimes. Although some deviations occur at low lifetimes, they remain beyond the sensitivity of current ATLAS searches. The second point is that consistently varying the LLP lifetimes and widths is, of course, theoretically sound, but in practice treating them as independent parameters does not significantly affect the results. This latter option is, on the other hand, computationally much more efficient, and indicates that the approach taken by the CMS and ATLAS collaborations is justified within the parameter ranges where they are sensitive.

%%%%%%%%%%%%%%%%%%%%%%%%%%%%%%%%%%%%%%%%%%%%%%%%%%%%%%%%%%%%%%%%%%%%%%%%%%%%%%%%%%%
\subsection{Going further: asymmetric production of LLPs}\label{sec:asymmetric}

\begin{figure}
    \centering
    \includegraphics[width=0.75\linewidth]{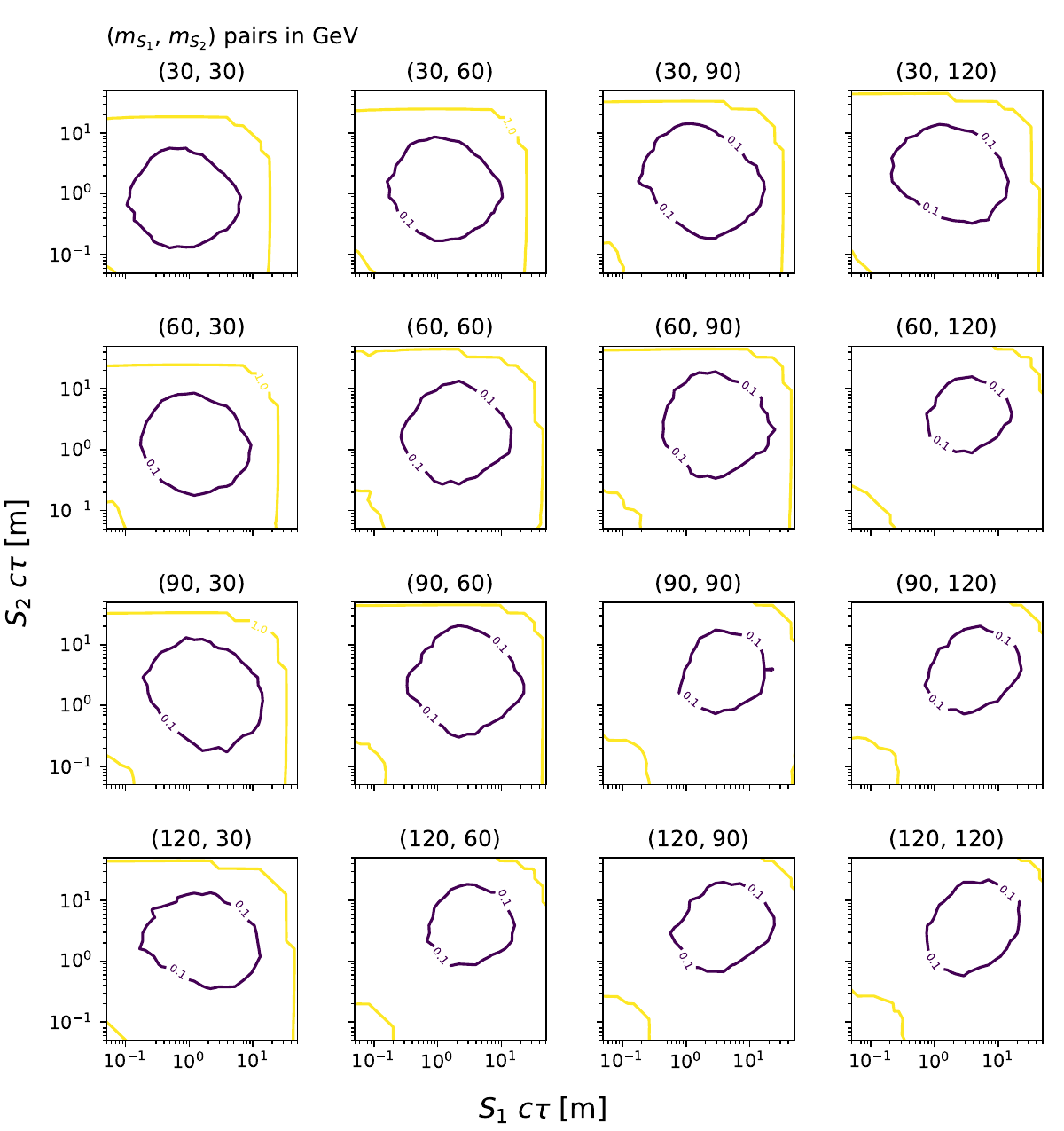}
    \caption{The observed 95\%CL exclusion on cross-section times branching fraction for an asymmetric model with a 200-GeV mediator, as a function of the two LLP decay lengths ($c\tau$, in meters) and masses in GeV (shown above each cell). The purple contour delimits regions excluded at 0.1~pb, and regions inside the yellow contour are excluded at the level of 1~pb.}
    \label{fig:mH200-Asym}
\end{figure}

\begin{figure}
    \centering
    \includegraphics[width=0.75\linewidth]{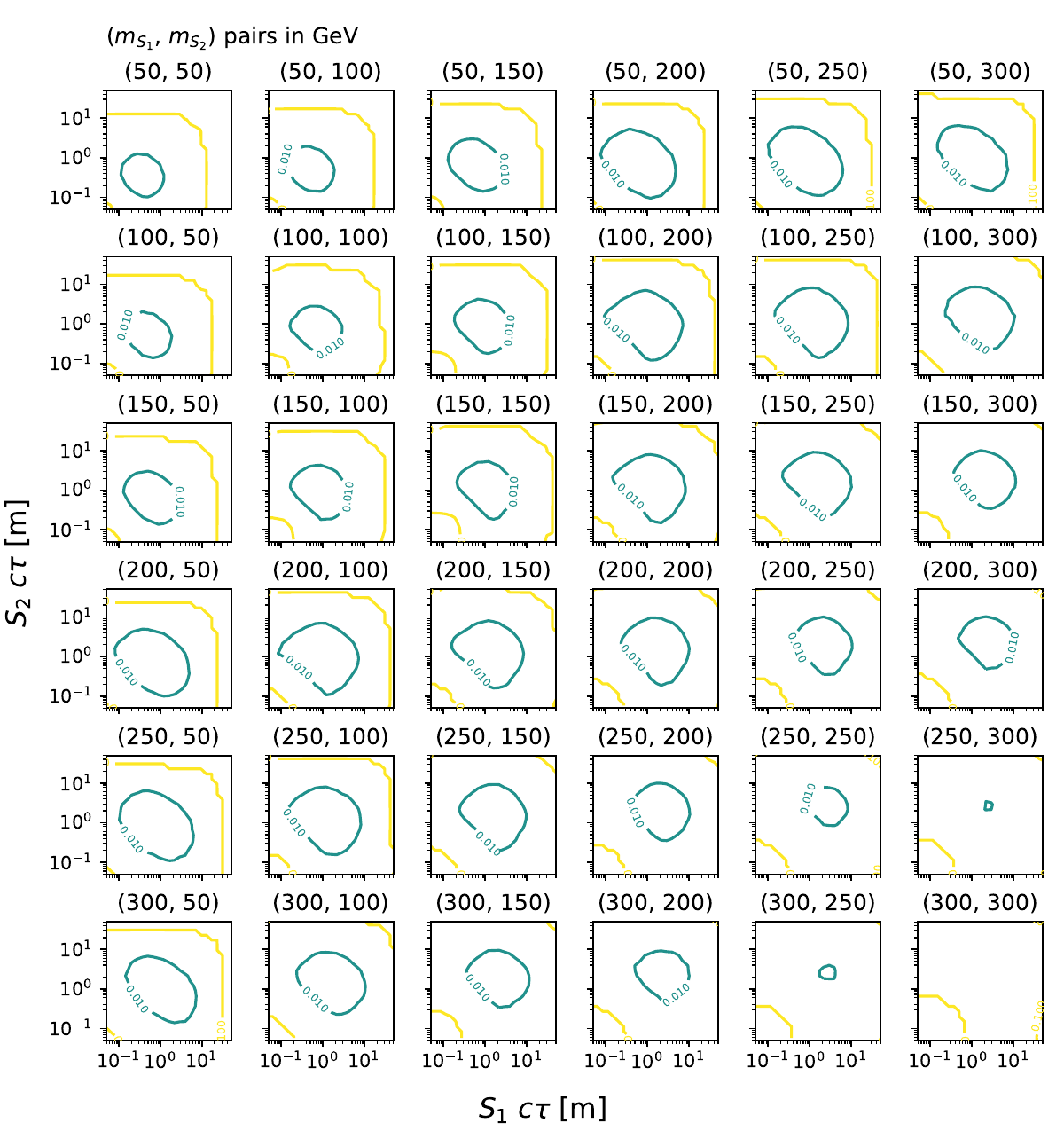}
    \caption{The observed 95\%CL  exclusion on cross-section times branching fraction for an asymmetric model with a 600-GeV mediator, as a function of the two LLP decay lengths ($c\tau$, in meters) and masses in GeV (shown above each cell). The cyan contour delimits regions excluded at 0.01~pb, and  regions inside the yellow contour are excluded at the level of 0.05~pb.}
    \label{fig:mH600-Asym}
\end{figure}

Having validated our theoretical and computational framework, we can now use it to go beyond the scenarios which are possible within the original HAHM template model. Namely, we now set the masses and lifetimes of the two LLPs to be different, and we use the surrogate models of Ref.~\cite{ATLAS-EXOT-2022-04} to extract efficiencies. Since, moreover, we have established that dissociating the LLP widths and lifetimes does not modify the results in any noticeable manner, we simply generate one set of events per $(m_{S_1}, m_{S_2})$ pair (choosing a macroscopic mean proper lifetime) and subsequently scan over $c\tau$ independently. The LLP mass pairs are chosen on a grid of possible values, which depend upon the mediator mass. At each point in the plane of the two LLP $c\tau$ values and for each point on the grid of LLP masses, we can obtain the selection efficiency for region A of Ref.~\cite{ATLAS-EXOT-2022-04} using the surrogate model. Moreover, the reference analysis published the observed and expected number of events in region A. We hence convert this efficiency value to an observed CL$_s$ exclusion value on the allowed cross-section times branching fraction, as shown in Figs.~\ref{fig:mH200-Asym} and~\ref{fig:mH600-Asym} for mediator masses of 200~GeV and 600~GeV respectively. This allows us to set limits on asymmetric LLP production for the first time.

For the 200-GeV case (Fig.~\ref{fig:mH200-Asym}) we see, as expected, that the 0.1~pb exclusion contours (violet) are broadly circular as a function of the logarithms of the two LLP lifetimes.  The looser exclusion contour, at 1~pb (yellow), has a ``corner pipe'' shape, dropping off when either LLP has a too-large lifetime (which implies a lack of reconstructible objects in the detector), or when both have too small lifetimes (leading to zero displaced trackless jets). The centroid of the exclusion contours shifts to higher lifetimes as the mass of the corresponding LLP increases, due to time dilation effects. Nevertheless, the excluded region at 0.1~pb corresponds to LLP lifetimes between 1~m and 10~m for most scenarios, which of course is the range where the calorimeter-based search is expected to be most sensitive.

The 600-GeV case (Fig.~\ref{fig:mH600-Asym}) exhibits similar features, but the exclusion contours are shifted to much lower values than in the 200-GeV case: 0.01~pb and 0.05~pb (cyan and yellow contours, respectively). The reference analysis is known to be more sensitive to heavy mediator scenarios, and correspondingly much lower cross-sections can be ruled out. Another interesting feature concerns the fact that the size of the 0.01~pb exclusion contour expands substantially as either of the LLP masses increases: this can be understood since heavier LLPs will produce more central jets, which are more efficiently selected by the analysis. If, however, the masses of \textit{both} LLPs increase (which happens around the diagonal of Fig.~\ref{fig:mH600-Asym}), eventually neither of them is produced with enough transverse momentum to be detected, and consequently the sensitivity almost vanishes.

%%%%%%%%%%%%%%%%%%%%%%%%%%%%%%%%%%%%%%%%%%%%%%%%%%%%%%%%%%%%%%%%%%%%%%%%%%%%%%%%%%%
%%%%%%%%%%%%%%%%%%%%%%%%%%%%%%%%%%%%%%%%%%%%%%%%%%%%%%%%%%%%%%%%%%%%%%%%%%%%%%%%%%%
%%%%%%%%%%%%%%%%%%%%%%%%%%%%%%%%%%%%%%%%%%%%%%%%%%%%%%%%%%%%%%%%%%%%%%%%%%%%%%%%%%%
\section{Conclusions}\label{sec:conclusions}

In this paper, we studied the constraints that can be set by searches for displaced jets at the Large Hadron Collider on the cross-section times branching fraction of ``asymmetric'' LLP production, i.e., the production of a pair of non-identical long-lived particles.  In passing, we commented upon a number of subtleties concerning the (unavoidable) usage of template models from the experimental collaborations and the applicability of existing constraints on different New Physics models.

We introduced a new toy model to study these asymmetric hadronic decays of long-lived scalars, addressing shortcomings of a commonly-used benchmark model. Our model, which allows the mediator to decay to two different long-lived scalars, uses LLP lifetimes as input parameters, rendering it more user-friendly. Using publicly available recasting material from a recent ATLAS search for displaced jets to calculate selection efficiencies, we validated that the lifetime-parametrised model in the symmetric limit yields consistent results with the existing benchmark models and that the lifetime can be treated as an independent parameter within the experimental search range. 
We then presented constraints on various combinations of LLP lifetimes and masses, for two mediator mass choices. Our findings show that existing searches can set constraints on the production cross-section times branching fraction of non-identical LLPs, at levels of 0.1 (0.01)~pb for a 200 (600) GeV mediator, for LLP lifetimes ranging from 10~cm to 10~m. Smaller mass configurations are more efficiently constrained at shorter LLP lifetimes, while heavier masses are better constrained at longer lifetimes. As either LLP becomes sufficiently short- or long-lived, the constraints weaken. 

%%%%%%%%%%%%%%%%%%%%%%%%%%%%%%%%%%%%%%%%%%%%%%%%%%%%%%%%%%%%%%%%%%%%%%%%%%%%%%%%%%%
%%%%%%%%%%%%%%%%%%%%%%%%%%%%%%%%%%%%%%%%%%%%%%%%%%%%%%%%%%%%%%%%%%%%%%%%%%%%%%%%%%%
%%%%%%%%%%%%%%%%%%%%%%%%%%%%%%%%%%%%%%%%%%%%%%%%%%%%%%%%%%%%%%%%%%%%%%%%%%%%%%%%%%%
\section*{Acknowledgments}\label{sec:Acknowledgments}

TC and LM would like to thank the Laboratoire de Physique de Clermont Auvergne for hospitality during the preparation of this paper. LC and AG are grateful to A. M. Teixeira for enlightening discussions.

%%%%%%%%%%%%%%%%%%%%%%%%%%%%%%%%%%%%%%%%%%%%%%%%%%%%%%%%%%%%%%%%%%%%%%%%%%%%%%%%%%%
%%%%%%%%%%%%%%%%%%%%%%%%%%%%%%%%%%%%%%%%%%%%%%%%%%%%%%%%%%%%%%%%%%%%%%%%%%%%%%%%%%%
%%%%%%%%%%%%%%%%%%%%%%%%%%%%%%%%%%%%%%%%%%%%%%%%%%%%%%%%%%%%%%%%%%%%%%%%%%%%%%%%%%%
\newpage

\bibliographystyle{JHEP}
\bibliography{references}
\end{document}